\documentclass[floatfix,twocolumn,showpacs,prl,aps,amsmath,superscriptaddress]{revtex4}
\usepackage{graphicx}
\usepackage{bm}
\usepackage{amssymb}
\usepackage{dcolumn}
\usepackage{color}

\begin{document}

\newcommand {\grbf}[1]{\rm{\boldmath $ #1 $}}
\newcommand{\vn}[1]{{\bf{#1}}}
\newcommand{\vht}[1]{{\boldsymbol{#1}}}
\newcommand{\matn}[1]{{\bf{#1}}}
\newcommand{\matnht}[1]{{\boldsymbol{#1}}}
\newcommand{\bege}{\begin{equation}}
\newcommand{\ee}{\end{equation}}
\newcommand{\bal}{\begin{aligned}}
\newcommand{\defbar}{\overline}
\newcommand{\SM}{\scriptstyle}
\newcommand{\eal}{\end{aligned}}
\newcommand{\udot}{\overset{.}{u}}
\newcommand{\exponential}[1]{{\exp(#1)}}
\newcommand{\phandot}[1]{\overset{\phantom{.}}{#1}}
\newcommand{\phandag}{\phantom{\dagger}}
\newcommand{\Trace}{\text{Tr}}
\def\sigic{\sigma^\q{int}}
\def\sigsj{\sigma^\q{sj}}
\def\sigij{\sigma^\q{int+sj}}

\newcommand{\ad}{\dagger}

\newcommand{\bra}[1]{\ensuremath{\langle #1  |}}
\newcommand{\ket}[1]{\ensuremath{| #1 \rangle}}

\newcommand{\matr}[1]{\left(\begin{matrix} #1  \end{matrix} \right)}

\newcommand{\q}[1]{\mathrm{#1}}
\newcommand{\mf}[1]{\ensuremath{\boldsymbol{\q{#1}}}}

\newcommand{\kf}{{\mf k}}
\newcommand{\qf}{{\mf q}}
\newcommand{\rf}{{\mf r}}
\newcommand{\Rf}{{\mf R}}
\newcommand{\xf}{{\mf x}}
\newcommand{\vf}{{\mf v}}
\newcommand{\pf}{{\mf p}}

\newcommand{\g}[1][]{\ensuremath{\hat{G}^{#1}}}
\newcommand{\gi}{\ensuremath{\slashed{\hat G}}}

\newcommand{\up}{\uparrow}
\newcommand{\down}{\downarrow}

\newcommand{\angles}[1]{\ensuremath{\left\langle #1 \right\rangle}}
\newcommand{\parens}[1]{\ensuremath{\left( #1 \right)}}
\newcommand{\bracks}[1]{\ensuremath{\left[ #1 \right]}}
\newcommand{\curls}[1]{\ensuremath{\left\{ #1 \right\}}}

\newcommand{\io}{\ensuremath{\q i\omega}}

\newcommand{\Romannumeral}[1]{\MakeUppercase{\romannumeral #1}}

\title{
\textbf{\emph{Ab Initio}} Theory of Scattering-Independent Anomalous Hall Effect}

\author{J\"urgen Weischenberg}
\affiliation{Peter Gr\"unberg Institut \& Institute for Advanced Simulation,
Forschungszentrum J\"ulich and JARA, 52425 J\"ulich, Germany}

\author{Frank Freimuth}
\affiliation{Peter Gr\"unberg Institut \& Institute for Advanced Simulation,
Forschungszentrum J\"ulich and JARA, 52425 J\"ulich, Germany}

\author{Jairo Sinova}
\affiliation{Department of Physics, Texas A\&M University, College Station,
Texas 77843-4242, USA}

\author{Stefan Bl\"ugel}
\affiliation{Peter Gr\"unberg Institut \& Institute for Advanced Simulation,
Forschungszentrum J\"ulich and JARA, 52425 J\"ulich, Germany}

\author{Yuriy Mokrousov}
\email[corresp.~author:~]{y.mokrousov@fz-juelich.de}
\affiliation{Peter Gr\"unberg Institut \& Institute for Advanced Simulation,
Forschungszentrum J\"ulich and JARA, 52425 J\"ulich, Germany}

\date{\today}

\begin{abstract}
We report on  first-principles calculations of the  side-jump contribution to the 
anomalous Hall conductivity  (AHC) directly from the electronic structure of a
perfect crystal. We implemented our approach for a short-range scattering 
disorder model within the density functional theory and computed the full 
scattering-independent AHC in elemental bcc Fe, hcp Co, fcc Ni, and L1$_0$ 
FePd and FePt alloys. The full AHC thus calculated agrees systematically with 
experiment to a degree unattainable so far, correctly capturing the previously 
missing elements of side-jump contributions, hence paving the way to a truly 
predictive theory of the anomalous Hall effect and turning it from a 
characterization tool to a probing tool of multi-band complex 
electronic band structures. 
\end{abstract}

\pacs{72.25.Ba, 72.15.Eb, 71.70.Ej}

\maketitle

The anomalous Hall effect (AHE) in ferromagnets is one of the most celebrated
transport phenomena in solid-state physics~\cite{hall}. It has been researched 
intensely in the past decade after it was realized that the intrinsic contribution  
(IC) could be interpreted in terms of the Berry phases of Bloch electrons in a 
solid~\cite{nagaosa06,nagaosa}. For almost ten years the IC was the only one, 
which could be accessed in density functional theory (DFT) calculations of the 
AHE~\cite{yao,wang,fermisurface}. The ability to estimate the impurity-driven 
(i.e.~extrinsic) contributions to the AHE has been remarkably limited so far, 
thus hindering the predictive power in understanding and engineering the AHE 
transport properties of real materials.

Besides the IC, $\sigma^\q{int}$, there are two extrinsic disorder-driven 
contributions to the anomalous Hall conductivity (AHC). In metallic systems 
they can be distinguished according to their parametric dependency on the impurity 
concentration $n$, with skew-scattering conductivity, $\sigma^\q{sk}$, proportional 
to $1/n$~\cite{smit}, and the {\it side-jump} contribution (SJC), $\sigma^\q{sj}$, 
independent of impurity concentration~\cite{berger}. The fact that the SJC, 
although originating from impurities, does not depend on their concentration, 
makes it one of the most challenging electron scattering mechanisms to understand 
and suggests a close relation to the intrinsic contribution of the AHE. This behavior 
arises in metallic systems from the $1/\epsilon_F\tau$ expansion 
of the transport coefficients in linear response for coupled multi-band systems. 
Consequently, since the SJC does not depend on the disorder strength, it cannot be 
easily separated from the IC in low temperature dc measurements~\cite{nagaosa}. 
So far, only for an L1$_0$-ordered FePd ferromagnetic alloy clear evidence has 
been presented that the side-jump contribution can dominate over other mechanisms 
of the AHE in a wide temperature range through a comparison of theoretical IC and 
experimentally measured AHC~\cite{seemann}. 
 
Previous numerical \textit{ab initio} treatments have mainly concentrated on the 
IC~\cite{nagaosa} and defined the SJC by taking the zero disorder limit of coherent 
potential approximation disordered alloys calculations~\cite{lowitzer}. Within such 
a treatment the indirect computation of the SJC presents a significant computational 
challenge considering also that the exact knowledge of the disorder potential in the 
system is necessary in this case. Since very often the experimental data are obtained 
on samples with unknown impurity content and disorder type, it is highly desirable 
to be able to evaluate the SJC explicitly from the electronic structure of a perfect crystal.  
A direct computation of the  SJC in models with disorder was not feasible until recently 
when it was shown that, assuming short-range uncorrelated disorder model, the SJC 
may be indeed calculated directly from the ideal electronic structure of a crystal without 
any disorder~\cite{sinova}, rendering possible a rigorous numerical study that can be 
fully compared to experimental results and can set the stage to a truly predictive 
theory of the AHE. While the validity of derived expressions has been demonstrated 
for simple models, values for the {\it scattering-independent} SJC in fundamental 
ferromagnetic materials such as Fe, Co or Ni, have not been obtained so far.  

In this Letter we report on calculations from  first principles of the values of the 
scattering-independent side-jump conductivity in elemental bcc Fe, hcp Co, 
fcc Ni, as well as ordered FePd and FePt alloys directly from the electronic 
structure of their pristine crystals. Our calculations unambiguously 
show that the calculated values of the total scattering-independent contributions, 
IC and SJC, agree systematically with experiments to a  level that was not reached 
before in these materials. More importantly, the inclusion of the scattering-independent 
SJC accounts consistently  for the discrepancy between the IC and the measured values 
in a very non-trivial fashion. We analyze the side-jump as a Fermi surface property and 
demonstrate that it shows a strong anisotropy with respect to the magnetization direction 
in the crystal, even more pronounced than that found for the intrinsic 
AHC~\cite{roman,hongbin}. Additionally, we demonstrate the importance of 
correlation effects in describing the AHE in fcc Ni correctly.

The starting point for the theory of the scattering-independent side-jump~\cite{sinova} 
are the retarded Green's function in equilibrium and the Hamiltonian $H$ of a general 
multiband noninteracting system in three spatial dimensions. At the first step we expand 
the self-energy of the system $\Sigma_\q{eq}$ in powers of potential $V(\rf)$, which 
describes scattering at impurities. For a short-range scattering disorder model, scalar 
delta-correlated Gaussian  disorder or delta-scattering uncorrelated disorder, the 
contribution to the self-energy which is of first order in $V(\rf)$ vanishes,  since one 
can assume that $\langle V(\rf)\rangle=0$ or else absorb $\langle V(\rf)\rangle$ into 
the Hamiltonian, a procedure which results in a simple shift of the energy levels. 
Further, inserting the expression for the self-energy within these simple disorder models 
into appropriate equations for the current densities derived following the Kubo-St\v{r}eda 
formalism, rotating into eigenstate representation and keeping only the leading order 
terms in the limit of vanishing disorder parameter $\mathcal V$,~i.e.~ignoring 
skew-scattering contributions, the scattering-independent part of the AHE conductivity 
may be written as 
$\sigma^{(0)}=\sigma^\q{int}+\sigma^\q{sj}$, where
\bege
\sigma^\q{int}_{ij}=\frac{e^2}{\hbar}\int\frac{d^3k}{(2\pi)^3}\q{Im}
{\sum\limits_{n\neq m} (f_n-f_m)\frac{
  v_{nm,i}(\kf)
  v_{mn,j}(\kf)}{(\omega_n-\omega_m)^2}}
  \label{ic}
\ee
can be recovered as the intrinsic contribution~\cite{yao}. In this expression  
indices $n$ and $m$ run over all bands with occupations $f_n$ and $f_m$, respectively, 
$v_{nm,i}$ are the matrix elements of the velocity operator 
$\hat v_{i}=\partial_{\hbar k_i}\hat H$, and $\omega_{n}(\kf)=\varepsilon_{n}(\kf)/\hbar$
with $\varepsilon_{n}(\kf)$ as band energies. The scattering-independent SJC to 
conductivity $\sigma^{(0)}$ reads for inversion-symmetric systems:
\bege  
\bal
&\sigma^{\q sj}_{ij}=\frac{e^2}{\hbar}\sum\limits_{n}
 \int\frac{d^3k}{(2\pi)^3}\q{Re}\,\q{Tr}\bigg\{
 \delta(\varepsilon_F-\varepsilon_{n})\frac{ \gamma_c}{[ \gamma_c]_{nn}}\times\\
 &\quad\times\bracks{ S_n  A_{k_i}( {1}-  S_n)
 \frac{\partial\varepsilon_n}{\partial k_j}
 -
 S_n A_{k_j}( {1}-  S_n)
 \frac{\partial\varepsilon_n}{\partial k_i}}\bigg\}.
\label{sjc}
\eal
\ee
Here $\varepsilon_F$ is the Fermi energy and the imaginary part of the self-energy 
$\q{Im}\Sigma_\q{eq}
=-\hbar\mathcal V\gamma$ is taken to be
in the eigenstate representation,~i.e.~$\gamma_c=U^\ad\gamma U$, with
\bege
{\gamma}=\frac 12 \sum\limits_{n}\int\frac{d^3 k}{(2\pi)^{2}}
\,U S_n U^\ad\,\delta(\varepsilon_F-\varepsilon_n),
\ee
$U$ as the $\mathbf{k}$-dependent unitary matrix that diagonalizes the Hamiltonian 
at point $\kf$,
\bege
[U^\ad H(\kf) U]_{nm}=\varepsilon_n(\kf)\delta_{nm},
\ee
$S_n$ is a matrix that is diagonal in the band indices, 
$[S_n]_{ij}=\delta_{ij}\delta_{in}$, and  the so-called Berry connection matrix is 
given by $ A_\kf=i U^\ad\partial_\kf  U $ \cite{sinova}. Not included in 
expression~(\ref{sjc}) are the vertex corrections,  which vanish for an inversion-symmetric 
system in the Gaussian disorder model. However, note that in contrast to the original 
formula as presented in Eq.~(3) of Ref.~\cite{sinova}, our expression for 
$\sigma^{\rm sj}_{ij}$ is manifestly antisymmetric. For the Rashba model it reduces 
to the original form of Eq.~(3) in Ref.~\cite{sinova}. It is important to 
note that the SJC in  the short-range disorder model, Eq.~(\ref{sjc}), is solely determined 
by the electronic structure of the pristine crystal and thus directly accessible by 
\textit{ab initio} methods. 

\begin{table}
\caption{\label{feco}
Anomalous Hall conductivities for bcc Fe and hcp Co in units of S/cm for selected 
high-symmetry orientations of the magnetization. $\sigic$, $\sigsj$ and $\sigij$ stand 
for IC, SJC and their sum, respectively. The experimental values are for the 
scattering-independent conductivity.}
\begin{ruledtabular}
\begin{tabular}{llll|lrr}
 {\bf Fe} & [001]  &  [111]   &  [110]   & {\bf Co}   &   $c$ axis  & $ab$ plane   \\ \hline
 $\sigic$ &  767   &   842    &   810    &  $\sigic$  &    477   & 100    \\
 $\sigsj$ &  111   &   178    &   141    &  $\sigsj$  &    217   & $-$30  \\
 $\sigij$ &  878   &  1020    &  951    &  $\sigij$  &    694   &  70     \\                      
 Exp.~\cite{dheer}     & 1032 &  &  &  Exp.~\cite{roman} &  813  &   150   
\end{tabular}
\end{ruledtabular}
\vspace{-0.5 cm}
\end{table}

In practice, we replace the integrals in Eqs.~(\ref{ic}) and~(\ref{sjc}) by a discrete sum 
over a finite number of $k$-points in the Brillouin zone (BZ).  To reduce the computational 
cost we adopt the method of Wannier interpolation~\cite{wang,yates}, which employs the 
description of the electronic structure in terms of maximally-localized Wannier functions 
(MLWFs), to evaluate Eqs.~(\ref{ic}) and~(\ref{sjc}) for bcc Fe, hcp Co, fcc Ni, as well as 
L1$_0$ FePd and FePt. The electronic structure calculations were performed with 
full-potential linearized augmented plane-wave method as implemented in the  J\"ulich 
DFT code \texttt{FLEUR}~\cite{fleur} within the generalized gradient approximation (GGA). 
We used the plane-wave cut-off $K_{\rm max}$ of 4.0 bohr$^{-1}$ and 16000 $k$-points 
for self-consistent calculations. Spin-orbit coupling was included in the calculations in 
second variation. We constructed a set of 18 MLWFs per atom using the \texttt{Wannier90} 
code~\cite{mostofi} and our interface between \texttt{FLEUR} and 
\texttt{Wannier90}~\cite{freimuth}. 

\begin{figure}[t!]
\begin{center}
\includegraphics[scale=0.38]{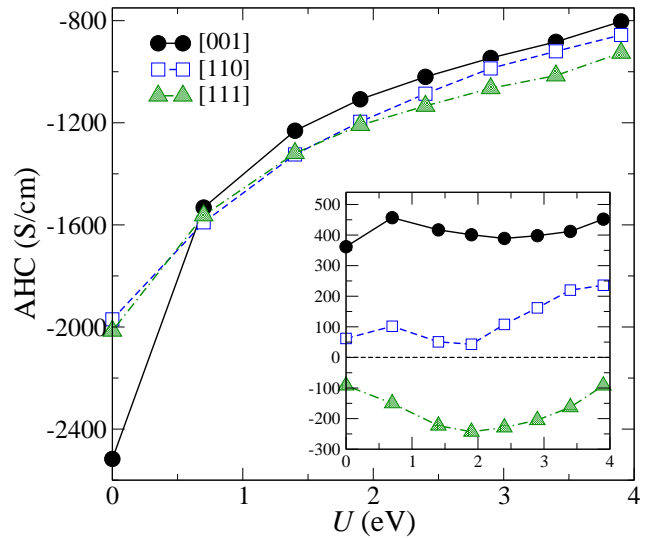}
\caption{\label{fig:Ni} (Color online) The dependence of the intrinsic and side-jump (inset) 
conductivities on the strength of the Coulomb repulsion $U$ of valence $d$-electrons in 
fcc Ni for different directions of the magnetization in the crystal.}
\end{center}
\end{figure}

We present the results of our calculations of the intrinsic and side-jump AHC for Fe, Co, 
Ni, FePd and FePt in Tables~I, II, and Fig.~\ref{fig:Ni} for high-symmetry directions of the 
magnetization $\mathbf{M}$ in the crystal. These results are compared  to experimental 
values, from which the skew scattering contribution was either explicitly 
substracted~\cite{fept-apl,seemann}, or can be safely ignored at higher 
temperatures~\cite{roman,dheer,miyasato}. 

We first analyze the results for bcc Fe. For $\mathbf{M}$ along [001], the IC in Fe accounts 
to roughly 75\% of the known experimental value of 
$\approx$1000~S/cm~\cite{fermisurface,dheer}. This comparison gets slightly improved 
considering that the experimental value averages over crystals with different orientation. 
Taking the SJC into consideration improves the value of the AHC in Fe significantly for all 
magnetization directions, with the angle-averaged $\sigic + \sigsj$ of about 90\% of the 
experimental conductivity. 
\begin{table}[b]
\vspace{-0.5 cm}
\caption{\label{feco}
Same as in Table I for L1$_0$ FePd and FePt alloys.}
\begin{ruledtabular}
\begin{tabular}{lll|lll}
 {\bf FePd}  &   [001]   &    [110]   &    {\bf FePt}   &    [001]    &   [110]       \\ \hline
 $\sigic$    &    133    &     280    &    $\sigic$     &     818     &    409        \\
 $\sigsj$    &    263    &     280    &    $\sigsj$     &     128     &    220        \\
 $\sigij$    &    396    &     560    &    $\sigij$     &     946     &    629        \\
 Exp.~\cite{seemann} & 806 & & Exp.~\cite{fept-apl,seemann} & $900\div 1267$ &  \\
\end{tabular}
\end{ruledtabular}
\end{table}
In hcp Co, the magnitude of the SJC for $\mathbf{M}$ along the $c$ axis is as large as
217~S/cm, with the total AHC of 694~S/cm, very close to the experimental value of 
about 800~S/cm. For $\mathbf{M}$ in the basal $ab$ plane the SJC is small and negative, 
bringing thus the intrinsic value down to $\approx$70~S/cm, somewhat away from the 
experimental value of about 150~S/cm. One has to keep in mind, however, that the 
experimental values for hcp Co are approximate~\cite{roman}.
 
A significant improvement upon including the SJC is also evident for the more complex 
ordered FePd and FePt alloys in their L1$_0$ phase with $\mathbf{M}$ along the [001] axis. 
For FePd, the IC is very small, of about 130~S/cm, while the side-jump AHC is twice as large 
and of the same sign, resulting in a value of the total AHC much closer to experiment, 
Table~II. As follows from our calculations,  in FePd the AHC is dominated by $\sigsj$, in 
accordance to an earlier indirect prediction~\cite{seemann}. On the other hand, in FePt 
with $\mathbf{M}\Vert$[001],  the IC is much larger, while the SJC is half the value 
of that in FePd. This is again in agreement to Ref.~\cite{seemann}, in which such a 
crossover between the intrinsic and side-jump conductivities, appearing within the Dirac 
model as well~\cite{sinitsyn}, was attributed to different SOI strength of Pd and Pt atoms. 
As far as the comparison to experiments is concerned, also in FePt adding the calculated 
SJC to the IC brings the total AHC within the range of experimentally observed values for 
samples of [001]-magnetized L1$_0$ FePt with high degree of ordering, $S>0.7$, and 
different sample thickness~\cite{seemann,fept-apl}.  
 
The case of fcc Ni presents a special challenge, since in this material the GGA value of 
the intrinsic AHC is much larger than the measured scattering-independent value, implying 
a sizable $\sigsj$ with the sign opposite to the IC~\cite{fermisurface}. From our calculations 
of the IC in fcc Ni we obtain  values which lie between $-$2000 and $-$2500~S/cm 
(Fig.~\ref{fig:Ni} at $U=0$), depending on the direction of $\mathbf{M}$, while 
the experimental value resides in the vicinity of $-$640~S/cm~\cite{lavine,fermisurface}. 
The calculated values for the scattering-independent SJC in fcc Ni presented in the same 
figure (at $U=0$) lie between $-$100 and 400~S/cm, and thus cannot explain the large 
discrepancy between theory and experiment. The description of the electronic structure 
of Ni within conventional DFT is well-known to be inaccurate, however. Several attempts 
aiming at improving the GGA values for quantities such as magnetocrystalline anisotropy 
energy, spin-wave dispersion~etc.,~were made in the past 
(e.g.~Refs.~\cite{kotliar},~\cite{ersoy} and references therein), proving the importance of 
correlation effects in this material and the sensitivity of calculated quantities on the shape 
of its Fermi surface.

In our work, we choose the GGA+$U$ approach in order to study the effect of correlations 
on the AHE in fcc Ni, following the implementation of Ref.~\cite{Shick} and treating the 
double counting corrections within the atomic limit~\cite{footnote2}. For this purpose, we 
scan the strength of intra-atomic repulsion parameter $U$, keeping at the same time the 
intra-atomic exchange parameter $J$ such that the value of the spin moment of Ni stays 
roughly constant~\cite{kotliar}. As can be seen from our calculations, presented in Fig.~1, 
the values of the $\sigic$ upon including $U$ change drastically and come closer to 
experiment, approaching a value of $-$800~S/cm when $U$ is changed in the range of 
0$-$4~eV, commonly used for calculations of other properties of Ni~\cite{ersoy,footnote}. 
This suggests that the main reason for the discrepancy between the intrinsic AHC values 
obtained from DFT and experiment might lie in the improper description of Ni's electronic 
structure from first principles. On the other hand, the values of the $\sigsj$ are affected 
differently by the modifications in the electronic structure, almost not changing for 
$\mathbf{M}$ along the [001] axis, and displaying a non-monotonous behavior within 
the range of 100--300 S/cm in the absolute value as a function of $U$ for two other 
magnetization directions.   

\begin{figure}[t!]
\includegraphics[scale=0.36]{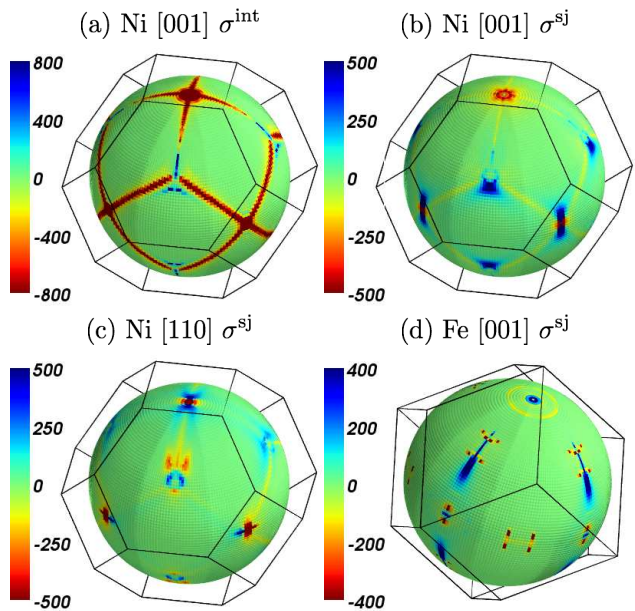}
\caption{\label{fig:sj} (Color online) 
Angle-resolved conductivity $d\sigma$/$d\Omega$ in 
units of S/cm as a function of direction in the BZ. 
$d\sigma$/$d\Omega$ corresponds to all contributions to $\sigma$ from inside 
the inner sphere in the BZ within the solid angle element $d\Omega$.
(a) $\sigma^{\text{int}}$ for Ni [001],
(b) $\sigma^{\text{sj}}$ for Ni [001], (c) $\sigma^{\text{sj}}$ 
for Ni [110], (d) $\sigma^{\text{sj}}$ for Fe [001].
}
\end{figure}

Such different sensitivity to the band structure can be understood by analyzing the 
structure of the SJC and the IC on the Fermi surface (FS). To simplify things, when 
taking into account the very complicated FSs of the ferromagnets considered in this 
study, we sum up the contributions to $\sigic$ and $\sigsj$ over all bands and all 
sheets of the FS, respectively, following Eqs.~(1) and (2), when going along a certain
direction in the BZ until the inner "Fermi" sphere in the BZ with the center at
its origin ($\Gamma$-point) is reached.  In the case of such angle-resolved IC in Ni
($U=0$), shown in Fig.~2(a), large contributions can be seen along the ''hot loops'' 
in the BZ, which are situated in the vicinity of the intersections between different 
sheets of bands~\cite{hongbin}, while the IC in the region away from such band 
crossings is also significant~\cite{fermisurface}. This is rather different from the 
topology of the SJC on the Fermi sphere. The SJC in Ni ($U=0$) and Fe, presented 
in Fig.~2(b)-(d) manifests that the main contribution to $\sigsj$ comes from certain 
isolated ''hot spots'', distributed rather sparsely over the Fermi sphere, while the SJC 
decays very quickly with the distance from such points.  Such a strong difference 
in the distribution of the $\sigsj$ and $\sigic$ on the FS arises from the effective 
magnetic monopole nature of the IC Berry's phase contribution near the band 
crossings, resulting in a more pronounced sensitivity of the $\sigic$ to the 
parameters of the electronic structure, such as Coulomb repulsion $U$, Fermi 
energy~etc., whereas the SJC does not contain such singularities near those 
crossings.

From our calculations presented above in Tables~I, II, Figs.~1 and 2, it is evident 
that $\sigsj$ exhibits large changes when the direction of the magnetization in 
the crystal is varied. In uniaxial crystals, such as FePt and hcp Co such anisotropy 
is not surprising, given that in these materials also the anisotropy of the intrinsic 
AHC appears already in the first order with respect to the directional cosines of 
the magnetization~\cite{roman}. And while the difference in the absolute change 
in the SJC and IC in FePd and FePt upon rotating the magnetization direction can 
be probably related to different SOI strength of the two materials~\cite{seemann}, 
in FePt the corresponding trend of the SJC and the IC is opposite owing to the 
different Fermi surface topology of the two contributions. Surprisingly, the strong
anisotropy of the SJC can be also observed in bcc Fe and fcc Ni. In Fe this anisotropy 
reaches as much as 70\%, while in Ni the SJC anisotropy is striking as compared to 
the IC anisotropy, with changes in sign and order of magnitude of $\sigsj$ as a 
function of the magnetization direction. This can be perhaps intuitively understood 
considering that $\sigsj$ is given almost entirely by singular ''hot spots'' at the FS, 
which change their position and the magnitude of their contribution depending 
on the matrix elements of the SOI, controlled in turn by the magnetization 
direction~\cite{roman,brataas}, compare~e.g.~Fig.~2(b) and (c) for Ni. 

Despite the unprecedented improvement of the values of the AHC in several 
ferromagnets when compared to the experimentally measured numbers, the 
scattering-independent SJC considered here cannot describe the entire physics 
of the complex side-jump scattering and will likely fail to describe it in certain 
systems where long-range scattering and spin-dependent scattering dominate. 
This can be particularly important for the case of low-doped, i.e., having long 
screening length, magnetic systems. Also, within our approach, we consider only 
the leading-order in impurity strength correction to the self-energy, which is 
justified within the weak scattering limit. This approximation might fail, however, 
when the perturbation in the crystal potential due to the presence of disorder or 
impurities is very strong. It would be highly desirable to extend the current model 
for the scattering-independent side-jump conductivity beyond the short-range 
disorder and weak scattering limit.

In summary, we have implemented a method to calculate the scattering-independent 
SJC within DFT. We found that for fundamental ferromagnets, such as Fe, Co, FePd 
and FePt the agreement between theory and experiment can be essentially improved 
upon considering the scattering-independent SJC. This SJC can be calculated from the 
electronic structure of the pristine crystal only, which encourages the application of 
the considered model for the side-jump scattering to wider classes of materials with 
the goal of extending the applicability of the DFT in treating complex transverse 
scattering phenomena and comparison to experiments that are commonly performed 
on samples with unknown disorder and impurity content.

We gratefully acknowledge J\"ulich Supercomputing Centre for computing time and 
funding under the HGF-YIG Programme VH-NG-513. JS was supported  
under Grant Nos. ONR-N000141110780 and NSF-DMR-1105512 and by the Research Corporation 
Cottrell Scholar Award.


\begin{thebibliography}{24}
\bibitem{footnote2} For several values of the parameter $U$ we compared our results to
the values of the AHC obtained in the "around mean field" limit, and did not find significant 
changes in the AHC.
\bibitem{footnote} For bcc Fe applying a GGA+$U$ approach with parameter $U$ 
of up to 2~eV results in a smooth positive shift of the values of the IC and SJC by 
up to approximately 60 and 40~S/cm, respectively, for all magnetization directions, 
bringing thus the total value of the calculated AHC even closer to experiment.
\end{thebibliography}
\end{document}